# Spatially-extended nonlinear generation of short-wavelength spin waves in YIG nanowaveguides

K. O. Nikolaev[1*#], B. Das Mohapatra[2#], G. Schmidt[2,3], S. O. Demokritov[1*], and V. E. Demidov[1]

[1]*Institute of Applied Physics, University of Muenster, 48149 Muenster, Germany*

[2]*Institut für Physik, Martin-Luther-Universität Halle-Wittenberg, 06120 Halle, Germany*

[3]*Interdisziplinäres Zentrum für Materialwissenschaften, Martin-Luther-Universität Halle-Wittenberg, 06120 Halle, Germany*

We experimentally study nonlinear propagation of spin waves in microscopic yttrium iron garnet waveguides, where the dispersion spectrum is engineered to enable efficient four-magnon interactions over a wide range of wavelengths. We show that under these conditions, the initial monochromatic spin wave nonlinearly generates co-propagating spin waves with well-defined, discrete frequencies. This process is characterized by a low energy threshold and can be observed in a wide range of frequencies and excitation powers. Thanks to the engineered dispersion, the process allows the generation of waves with short wavelengths that cannot be excited directly by a linear excitation mechanism. The nonlinearly generated short-wavelength spin waves continuously acquire the energy from the initial pump wave during co-propagation, which results in compensation of their propagation losses over significant distances. The observed phenomena can be used to implement frequency- and wavelength-conversion operations in magnonic nanodevices and circuits.

[#]These authors contributed equally

*Corresponding author, e-mail: k.nikolaev@uni-muenster.de

# I. INTRODUCTION

Nonlinear spin-wave phenomena are widely considered within the field of magnonics for the implementation of various signal-processing and computing tasks [1-6]. The nonlinearity of spin waves arises from the intrinsic nonlinearity of the equation of motion for the magnetization, which leads to a strong dependence of the propagation characteristics of spin waves on their amplitude [7]. In particular, this nonlinearity causes modification of the dispersion spectrum of spin waves with the change in their amplitude, which can be used to implement nonlinear phase shifting, modulation, interference, and multiplexing [4,8-13], as well as the nonlinear transformation of the wavelength of spin waves [14,15]. Another important class of nonlinear spin-wave phenomena is constituted by the phenomena associated with the nonlinear coupling between spin waves with different frequencies and different wavelengths, which can be described in terms of multi-magnon splitting and confluence processes [7]. On the one hand, these processes often play a negative role. For example, they are known to increase the attenuation of propagating spin waves and limit their amplitudes due to unwanted energy transfer to parasitic spin-wave modes [16-20]. On the other hand, by controlling nonlinear energy flows, it becomes possible to implement various frequency-conversion operations [21-28] which can be used for advanced information processing.

It is important to emphasize that most multi-magnon processes require the amplitude of the interacting spin waves to exceed a certain threshold determined by natural magnetic damping [7]. Therefore, in conductive ferromagnets where the damping is relatively strong, it is often necessary to use additional damping-compensation mechanisms to enable nonlinear interactions [27,29], which otherwise cannot be observed at reasonable excitation powers. In this regard, the use of low-damping magnetic insulators, such as yttrium iron garnet (YIG) [30-32], offers rich opportunities for studies of nonlinear magnon interactions and their technical applications. Owing to recent developments in the preparation of high-quality, nanometer-thick films, YIG can now be structured at the sub-micrometer scale [33,34]. This structuring not only provides the opportunity to implement nano-scale spin-wave devices, but also allows one to efficiently control the dispersion spectrum of spin waves through confinement effects. In turn, the deterministic control over the magnon spectrum through such structuring allows for the suppression of detrimental nonlinear interactions while enhancing the desired effects [24, 28]. Indeed, for the emergence of nonlinear magnon interactions, not only strong nonlinearity is required, but also the overall energy and momentum of interacting magnons must be conserved, which can be either forbidden or enabled depending on their dispersion. Therefore, the possibility to engineer the dispersion spectrum is essential for the development of magnonic devices based on nonlinear phenomena.

Here we show that by adjusting the geometrical parameters of YIG nanowaveguides, one can precisely balance the contributions of dipolar and exchange interactions and obtain a nearly linear dispersion relation for spin waves over a wide range of wavelengths. Under these conditions, the energy and momentum conservation rules can be easily satisfied, which results in cascaded four-magnon processes, leading to the efficient energy transfer from the initially excited long-wavelength primary monochromatic spin wave to short-wavelength spin waves with controlled frequencies. In contrast to the four-magnon process previously observed in waveguides based on conductive ferromagnets [27], in YIG waveguides with a tailored dispersion spectrum, the process does not require the use of additional damping compensation mechanisms and allows the nonlinear generation of spin waves with wavelengths down to 350 nm. Due to the low intrinsic damping and highly efficient nonlinear coupling, the process also allows the complete compensation of the spatial attenuation of nonlinearly generated waves at propagation distances up to 20 µm. This highly efficient process has large potential for technical applications, since it allows the controllable conversion of frequencies and wavelengths of propagating spin waves.

## II. EXPERIMENT

Figure 1 shows the measurement schematic of the experiment, wherein we study the propagation of spin waves in YIG waveguides with a width of 500 nm and thickness of 45 nm. The waveguides are fabricated using electron-beam lithography and lift-off technique from a YIG film deposited by pulsed-laser deposition [32]. As determined from independent measurements, the YIG film is characterized by a saturation magnetization $\mu_0 M_s = 162$ mT and a Gilbert damping parameter $\alpha = 4\times10^{-4}$. The waveguide is magnetized by a static magnetic field $H_0$ applied perpendicular to its axis. Spin waves are inductively excited by a Au antenna with a thickness of 200 nm and a width of 500 nm by applying a monochromatic microwave current at frequency $f$. The power of the excitation microwave signal is varied in the range $P = 0.01$-$1.5$ mW. Propagating spin waves are detected using micro-focus Brillouin light scattering (BLS) spectroscopy [35]. The probing laser light with a wavelength of 473 nm and a power of 0.25 mW is focused on the surface of the waveguide into a diffraction-limited spot (see Fig. 1) using a microscope objective lens with a magnification of 100 and a numerical aperture of 0.9. Due to the interaction of the probing light with spin waves, it becomes modulated. The intensity of modulation at a given frequency (BLS intensity) is proportional to the intensity of the spin wave at this frequency, which allows recording the frequency spectra of spin waves. By moving the probing spot relative to the sample surface, the spin-wave spectra at different spatial positions can be measured. This allows one to analyze the process of nonlinear generation of new spin waves with different frequencies with a high spatial resolution.

Before moving to the experimental results, we first discuss the choice of parameters used in our experiments. As mentioned above, the peculiarities of the dispersion spectrum of spin waves are essential for achieving efficient nonlinear interactions in a wide range of wavelengths. Figure 2(a) shows the experimentally-measured and numerically-calculated dispersion curve of spin waves in the used 45-nm thick YIG waveguide, along with numerically-calculated curves obtained for thicknesses $t$ = 20 and 100 nm. The dispersion curves are calculated at $\mu_0 H_0$=50 mT using the micromagnetic simulation package mumax3 [36] and the approach described in detail in Ref. [37]. The standard value for the YIG exchange constant 3.66 pJ/m is used in the calculations. All other parameters are set in accordance with the values known from the experiment. To prove the validity of the calculations, spin waves propagating in the studied 45-nm thick YIG waveguide were mapped using phase-resolved BLS measurements, which allow direct determination of the wavelength of spin waves at a given frequency [35]. The results of these measurement (unfilled diamonds in Fig. 2(a)) are in excellent agreement with the results of numerical simulations. We note that the frequency range in phase-resolved measurements is limited from above by $f_{max}$ ≈ 3.1 GHz, which corresponds to the wavelength of spin waves about 700 nm. At higher frequencies, the efficiency of linear excitation of spin waves by a 500-nm wide antenna decreases drastically, which does not allow one to directly measure the dispersion curve for wavenumbers $k$ > 9 µm$^{-1}$. However, good agreement between experimental data and results of simulations within the range 0 < $k$ < 9 µm$^{-1}$ allows one to rely on data obtained from calculations for $k$ > 9 µm$^{-1}$.

To better illustrate the effects of the thickness of the waveguide on the spin-wave dispersion, we show in Fig. 2(b) the $k$-dependences of the group velocity approximated as $v_g$ ≈ 2π$\Delta f$/$\Delta k$ obtained from the data in Fig. 2(a). As seen from these data, in waveguides with the thickness $t$ = 20 nm and 100 nm, the group velocity changes strongly over the entire range of $k$ from 0 to 20 µm$^{-1}$. On the contrary, in the case of $t$ = 45 nm, chosen for our experiments, $v_g$ is almost constant in a wide range of 2 < $k$ < 20 µm$^{-1}$, which corresponds to a nearly linear dispersion curve in this range (Fig. 2(a)). This is due to the precise balance of the contributions of dipolar and exchange interactions to the group velocity of spin waves achieved at $t$ = 45 nm. At small film thicknesses (see the data for $t$ = 20 nm in Fig. 2(b)), the dominant exchange interaction leads to a monotonic increase of the group velocity with increasing $k$. At large thicknesses (see the data for $t = 100$ nm in Fig. 2(b)), $v_g$ exhibits a strong increase at small $k$ caused by the influence of the dipole interaction [38]. As seen from Fig. 2(b), these two effects nearly cancel each other at $t = 45$ nm in a broad range of $k$.

The most obvious application-relevant consequence of the constant group velocity is the dispersionless propagation of short spin-wave pulses in a waveguide with the optimized thickness $t = 45$ nm [38]. More importantly, the fact that the slope of the dispersion curve $\Delta f/\Delta k$ is effectively constant enables fulfillment of the energy and momentum conservation conditions for non-degenerate nonlinear four-magnon interaction processes over a wide range of $k$. In such processes, two primary magnons with a frequency $f_m$ and a wavenumber $k_m$ create a pair of secondary magnons, with frequencies $f_m \pm \delta f$ and wavenumbers $k_m \pm \delta k$. In order for this process to be efficient, the dispersion curve must have a constant slope below and above the initial wavenumber $k_m$. Otherwise, the process is allowed only in a small vicinity of $k_m$, where any dispersion curve can be considered as approximately linear [27]. As will be shown below, the linearity of the dispersion relation for the YIG waveguides used in our experiments indeed enables observation of four-magnon processes in a very wide $k$-range.

## III. RESULTS AND DISCUSSION

Figure 3 illustrates the four-magnon interaction process observed in the studied YIG waveguides. In this experiment, we apply to the antenna a monochromatic excitation signal at a frequency $f_{exc}$ and power $P$ increasing from 0.01 mW to 1.5 mW, and measure the frequency spectra of spin waves in the waveguide at a distance $y = 10$ μm from the antenna. In Figs. 3(a)-3(c), we present the obtained spectra in the form of color maps that show the BLS intensity in frequency-power coordinates. Figures 3(a)-3(c) correspond to the frequencies of the excitation signal $f_{exc}$= 2.7, 2.8, and 2.9 GHz, respectively. Figures 3(d)-3(f) show representative spectra corresponding to the sections of the color maps at the labeled values of the excitation power.

As expected for the linear propagation regime, at the relatively low power of 0.1 mW, all spectra show a single peak centered at the excitation frequency (Figs. 3(d)-(f)). As the power increases above a certain threshold, new peaks appear at discrete frequencies below and above $f_{exc}$. This indicates an onset of nonlinear energy transfer from the primary monochromatic spin wave to certain discrete spin-wave modes, leading to the generation of new discrete spectral components. At $f_{exc}$ = 2.7 GHz (Fig. 3(a)), this discrete-frequency generation regime is observed only in a small power window $P$ = 0.3-0.5 mW. At higher powers, it is followed by the formation of an almost continuous spectrum of nonlinearly generated spin waves. However, as the excitation frequency increases (Figs. 3(b), 3(c)), the discrete-frequency generation regime becomes more stable. At $f_{exc}$= 2.8 GHz (Fig. 3(b)), it is observed up to $P$ = 1.2 mW, and at $f_{exc}$= 2.9 GHz (Fig. 3(c)), it extends up to the maximum power used in the experiment $P$ = 1.5 mW.

To characterize in more detail the onset of the nonlinear spin-wave generation, we show in Fig. 4(a) the power dependence of the BLS intensity for the primary wave (vertical section of the map in Fig. 3(c) at the excitation frequency 2.9 GHz). As seen from these data, the intensity of the primary wave increases linearly with the increase in $P$ up to 0.5 mW. This indicates that up to this power, the system remains in the linear regime. At $P$ > 0.5 mW, the intensity drops abruptly and new frequency components appear in the spectrum (Fig. 3(c)). This leads us to the conclusion that the observed nonlinear generation is the dominating nonlinear phenomenon in the studied system, which is not preceded by any other nonlinear scattering/damping phenomena at lower powers. The power at which the drop of the intensity occurs can be considered as the threshold power, $P_{th}$, for the observed process (Fig. 4(a)). As seen from Fig. 4(b), $P_{th}$ increases with the increase in the excitation frequency and becomes larger than the maximum used power as the frequency approaches $f_{max}$ = 3.1 GHz. We attribute this to the decrease in the excitation efficiency of the primary spin wave by the used antenna, as discussed above.

We now discuss the physical origin of the nonlinearly generated discrete frequency components. As seen from Figs. 3(d)-3(f), the new peaks form an equidistant spectrum and the frequency separation between the peaks increases with the increase of the excitation frequency $f_{exc}$. However, regardless of $f_{exc}$, the spectral peak appearing below $f_{exc}$ is always located at the same frequency $f^*$, which is very close to the frequency of spin waves with a wavenumber $k = 0$ (Fig. 2(a)). The frequency of the first peak appearing above $f_{exc}$ is $f_1 = f_{exc} + \delta f$, where $\delta f = f_{exc} - f^*$, and the frequency of the second peak is $f_2 = f_{exc} + 2\delta f$. These behaviors are characteristic of the four-magnon interaction process. Figure 5 illustrates this process for the case of $f_{exc}$ = 2.9 GHz. Here, we plot the dispersion curve obtained from micromagnetic simulations (squares) and mark by circles the spin-wave states corresponding to the frequencies of the peaks observed in the experiment (Fig. 3(f)). As seen from these data, due to the constant slope of the dispersion curve, all four circles are located on a straight line and are equidistant in both frequency and wavenumber. This corresponds to the fulfillment of the energy and momentum conservation conditions for four-magnon processes illustrated by arrows in Fig. 5. In particular, two initially excited magnons at $f_{exc}$ can create a pair of magnons at $f^*$ and $f_1$, and two magnons at $f_1$ can create a pair of magnons at $f_{exc}$ and $f_2$.

Note that the constant slope of the dispersion curve results in the fulfillment of the conservation conditions for four-magnon processes with arbitrary $\delta f$, which can lead to the population of all magnon states on the linear part of the dispersion curve. Indeed, this is what we observe in the experiment at high excitation powers (see, e.g., Fig. 3(a), $P > 0.5$ mW). However, the experimental results clearly show that the process where $\delta f = f_{exc} - f^*$, is more energetically favorable. We associate this with the small group velocities of magnons in the vicinity of $k = 0$ (Fig. 2(b)). As already mentioned, the power threshold of the four-magnon process is determined by the damping rate. In turn, this rate is determined by the intrinsic damping and losses due to energy transport by spin waves. The latter contribution decreases with the decrease in the group velocity and, therefore, minimizes for magnons in the vicinity of $k = 0$.

We now characterize the nonlinearly generated spin waves at frequencies $f_1$ and $f_2 > f_{exc}$. Using the calculated dispersion curve (Fig. 5), we determine the corresponding wavenumbers $k_1 = 12.1$ $\mu m^{-1}$, $k_2 = 17.8$ $\mu m^{-1}$ and wavelengths $\lambda_1 = 520$ nm, $\lambda_2 = 350$ nm. We emphasize that these wavelengths are significantly shorter than the wavelength of 700 nm, corresponding to the frequency $f_{max} = 3.1$ GHz (Fig. 5), at which the linear excitation mechanism becomes completely inefficient. In other words, the observed nonlinear process allows the generation of short-wavelength spin waves that cannot be excited directly by the inductive antenna.

Figure 6 characterizes the propagation of the generated spin waves. Here, we show the spatial dependences of the maximum intensity in the spectral peaks at frequencies $f_1$ (Fig. 6(b)) and $f_2$ (Fig. 6(c)) along with the dependence for the initial wave at $f_{exc}$ (Fig. 6(a)). As seen from these data, the intensity of the wave at $f_1$ (Fig. 6(b)) increases quickly with distance from the antenna ($y = 0$ corresponds to the center of the antenna). It reaches a maximum at $y = 2$ $\mu m$ and then remains approximately constant up to $y = 9$ $\mu m$. This indicates that energy transfer from the primary wave at $f_{exc}$ to the wave at $f_1$ develops in the immediate vicinity of the excitation point. This transfer strongly affects the intensity of the primary wave (squares in Fig. 6(a)): in the interval $y = 1$-9 $\mu m$ the intensity at $f_{exc}$ decreases by about 60%. Note that this decrease is much larger than the decrease due to linear Gilbert damping, which is seen from the linear-regime spatial dependence measured at a small excitation power of 0.01 mW (triangles in Fig. 6(a)). This indicates that a large part of the energy of the primary wave is transferred to the wave at $f_1$ in the initial propagation stage. As seen from Fig. 6(b), in the interval $y = 1$-9 $\mu m$, this transfer completely compensates for energy losses of the wave at $f_1$. Note that the wave at $f_1$ loses energy not only due to the linear damping, but also due to nonlinear energy transfer to the wave at $f_2$. The latter process is likely responsible for the fast attenuation of the intensity at $f_1$ in the interval $y$ = 10-20 µm. As seen from Fig. 6(c), the energy transfer to the wave at $f_2$ develops on a noticeably larger spatial scale: the intensity of the wave at $f_2$ gradually increases in the interval $y$ = 1-5 µm. At larger distances $y$ = 5-20 µm, the intensity remains almost constant, which indicates that the nonlinear energy flow into the wave at $f_2$ completely compensates for its linear attenuation over this spatial interval. Finally, at $y > 20$ µm, the intensity at $f_2$ begins to decrease in space. Simultaneously, the spatial attenuation of the wave at $f_1$ becomes much weaker. Most likely, this indicates that the nonlinear energy flow from the wave at $f_1$ to the wave at $f_2$ stops due to a decrease in the intensity of the former wave below the interaction threshold. We emphasize that our observations show that the four-magnon process can be used not only to generate spin waves, but also to realize their decay-free propagation over long distances, which previously could only be achieved using sophisticated damping-compensation mechanisms [37].

## IV. CONCLUSIONS

In this work, we have shown that spin-wave nanowaveguides fabricated from nanometer-thick YIG films provide unique opportunities for engineering the dispersion spectrum of spin waves, which is of great importance for the ability to control nonlinear spin-wave interactions and use them for signal processing in magnonic devices and circuits. In particular, by tuning the geometrical parameters of nanowaveguides, it becomes possible to achieve a nearly linear dispersion relation for spin waves, which leads to the fulfilment of the conservation conditions for multi-magnon nonlinear interactions in a wide range of wavelengths. As a result, controllable nonlinear conversion of frequency and wavelength of spin waves can be realized, which allows efficient generation of spin waves with short wavelengths and spatially-extended compensation of their propagation losses. Additionally, the broadband linearity of the dispersion relation enables synchronous propagation and nonlinear interaction of spin-wave pulses with different frequencies, which is important for the development of high-data-rate nonlinear magnonic devices operating with ultra-short spin-wave pulses.

## ACKNOWLEDGMENTS

This work was supported by the Deutsche Forschungsgemeinschaft (DFG, German Research Foundation) – project number 529812702. We would also like to thank Stephanie Lake and Anoop Kamalasanan for their help during the fabrication of the sample and Seth Kurfman for useful suggestions on the manuscript.

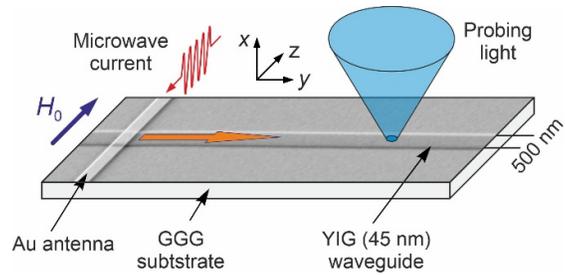

FIG. 1. Schematics of the experiment. Spin waves in YIG waveguides with a width of 500 nm and a thickness of 45 nm are excited using a 200-nm thick and 500-nm wide Au antenna. The structure is magnetized by a static magnetic field $H_0$ applied perpendicular to the waveguide axis. Frequency spectra of spin waves are measured with spatial resolution using micro-focus BLS spectroscopy.

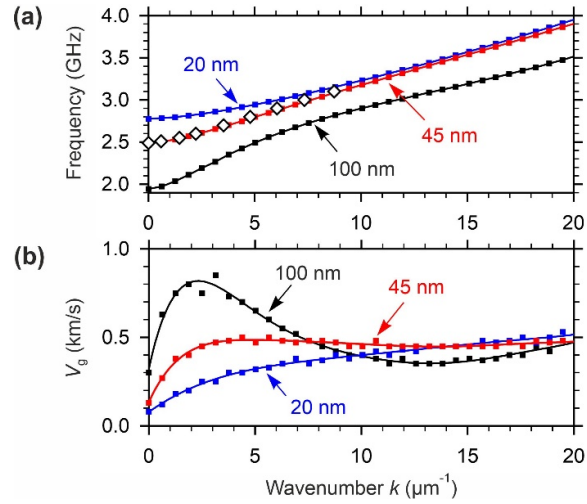

FIG. 2. (a) Dispersion curve of spin waves in the used 45-nm thick YIG waveguide along with the curves obtained for thicknesses $t = 20$ and 100 nm, as labeled. Filled squares show results of micromagnetic simulations. Unfilled diamonds show experimental data. Curves are guides for the eye. (b) $k$-dependences of the group velocity $v_g$ for different thicknesses, as labeled. The data are obtained at $\mu_0 H_0 = 50$ mT.

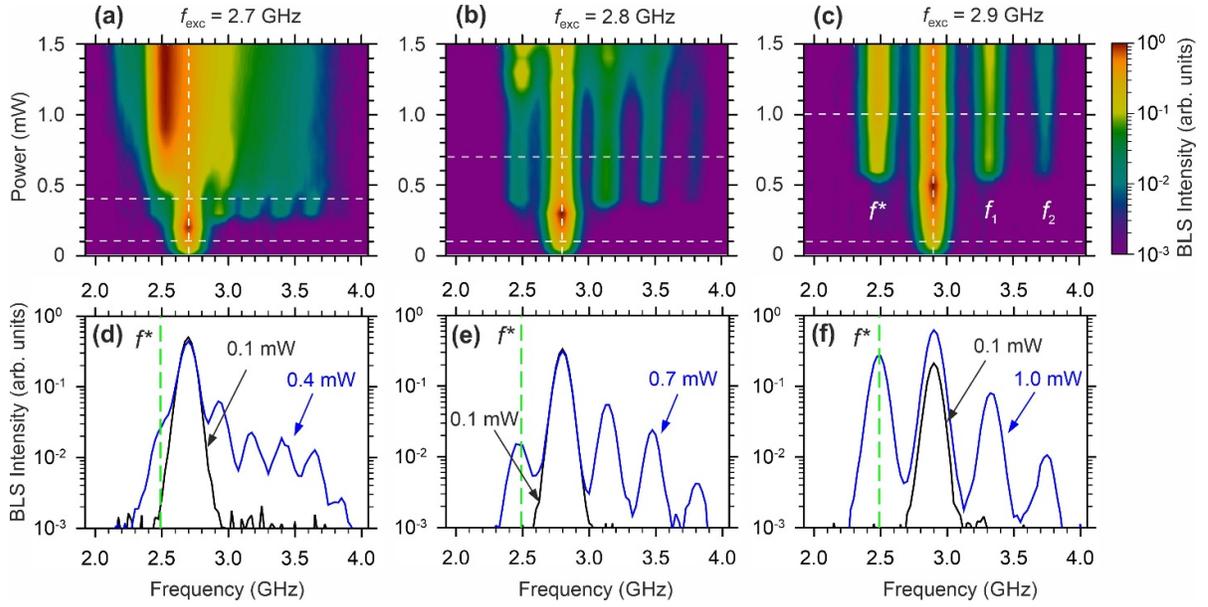

FIG. 3. (a)-(c) Spin-wave spectra in the form of color maps of BLS intensity in frequency-power coordinates, measured at frequencies of the excitation signal $f_{exc}$ = 2.7, 2.8, and 2.9 GHz, respectively. Vertical dashed lines mark $f_{exc}$. (d)-(f) Representative spectra corresponding to horizontal sections of the color maps at the labeled values of the power (horizontal dashed lines in (a)-(c)). Vertical dashed lines indicate the nonlinearly generated mode with lowest frequency $f^*$. The data are obtained at $\mu_0 H_0$=50 mT.

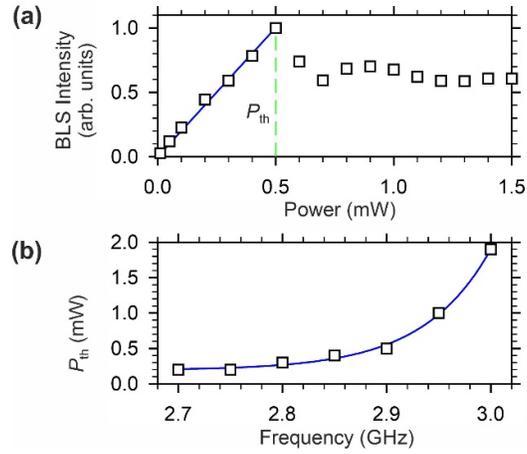

FIG. 4. (a) Power dependence of the BLS intensity at the excitation frequency $f_{exc}=2.9$ GHz. Squares show experimental data. Line shows the linear fit of the data at $P < 0.5$ mW. Vertical dashed line marks the threshold power $P_{th}$ for the observed nonlinear process. (b) Frequency dependence of the threshold power. Squares show experimental data. Curve is a guide for the eye. The data are obtained at $\mu_0 H_0 = 50$ mT.

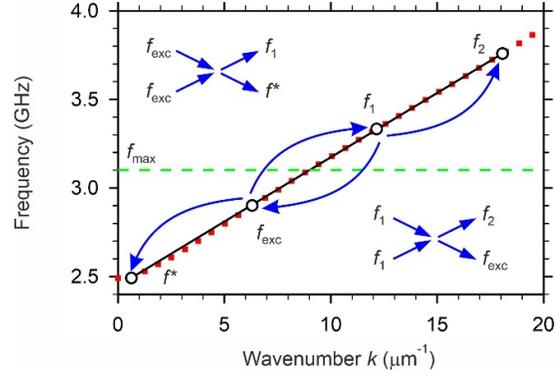

FIG. 5. Dispersion curve for spin waves in the studied YIG waveguide obtained from micromagnetic simulations (squares). Circles mark the spin-wave states corresponding to the frequencies of the peaks observed in the experiment at $f_{exc}$= 2.9 GHz. Note that all four circles are located on a straight line. Arrows illustrate four-magnon processes leading to the generation of new discrete spectral components. Horizontal dashed line marks the frequency $f_{max}$, at which the linear excitation mechanism becomes completely inefficient.

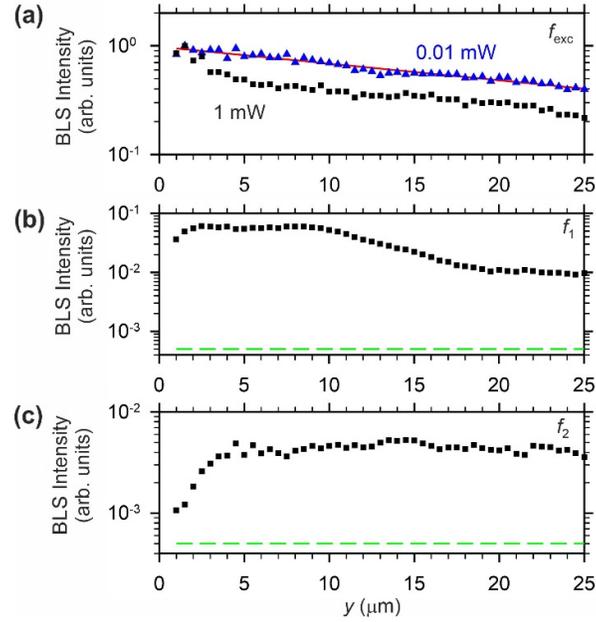

FIG. 6. Spatial dependences of the intensity in the spectral peaks at frequencies $f_{exc}$ (a), $f_1$ (b), and $f_2$ (c). Squares show experimental data obtained at $P$ = 1 mW. For comparison, triangles in (a) show the spatial dependence of the intensity of the initial wave measured at low excitation power $P$ = 0.01 mW. Solid line in (a) shows the exponential fit of experimental data at $P$ = 0.01 mW. Horizontal dashed lines in (b) and (c) indicate the noise level. $y$ = 0 corresponds to the center of the antenna. The data are obtained at $\mu_0 H_0$=50 mT and $f_{exc}$= 2.9 GHz.